\documentclass[prl,aps,showpacs,floatfix,floats,superscriptaddress,nobalancelastpage,twocolumn]{revtex4-1}
\usepackage{graphicx}
\usepackage{graphics}
\usepackage{latexsym,amsmath,amssymb,bm,euscript}
\usepackage{color}
\usepackage{dcolumn}
\usepackage{bm}
\usepackage{epstopdf}
\usepackage{times}
\usepackage{multirow}
\usepackage{wasysym}
\usepackage{subfigure}
\usepackage{bbm}

\def\ra{\rangle}
\def\la{\langle}

\begin{document}
\title{Antiferroquadrupolar order and rotational symmetry breaking in a generalized bilinear-biquadratic model on a square lattice} 
\author{Hsin-Hua Lai}
\affiliation{Department of Physics and Astronomy \& Rice Center for Quantum Materials, Rice University, Houston, Texas 77005, USA}
\author{Wen-Jun Hu}
\affiliation{Department of Physics and Astronomy \& Rice Center for Quantum Materials, Rice University, Houston, Texas 77005, USA}
\author{Emilian M. Nica}
\affiliation{Department of Physics and Astronomy \& Rice Center for Quantum Materials, Rice University, Houston, Texas 77005, USA}
\author{Rong Yu}
\affiliation{Department of Physics, Renmin University of China, Beijing, 100872, China}
\affiliation{Department of Physics and Astronomy, Shanghai Jiao Tong
University, Shanghai 200240, China and Collaborative Innovation
Center of Advanced Microstructures, Nanjing 210093, China}
\author{Qimiao Si}
\affiliation{Department of Physics and Astronomy \& Rice Center for Quantum Materials, Rice University, Houston, Texas 77005, USA}
\date{\today}

\begin{abstract}
The magnetic and nematic properties of the iron chalcogenides have recently been the subject of intense interest. Motivated by the proposed antiferroquadrupolar and Ising-nematic orders for the bulk FeSe, we study the phase diagram of an $S=1$ generalized bilinear-biquadratic model with multi-neighbor interactions. We find a large parameter regime for a ($\pi$,0) antiferroquadrupolar phase, showing how quantum fluctuations stabilize it by lifting  an infinite degeneracy of certain semiclassical states. Evidence for this C$_4$-symmetry-breaking quadrupolar phase is also provided by an unbiased density matrix renormalization group analysis. We discuss the implications of our results for FeSe and related iron-based superconductors.
\end{abstract}
\maketitle

\textit{Introduction}---
Much of the current effort in the study of the iron-based superconductors (FeSCs) is devoted to understanding the magnetism in their normal state \cite{Kamihara08, Si2016}. While the iron pnictides were the focus of the early effort in the FeSC field, iron chalcogenides have occupied the center stage more recently.
Among them, FeSe takes a special place.  In the single-layer limit, FeSe has the highest superconducting transition temperature among the FeSCs~\cite{QYWang2012,JJLee_Nature,SLHe2013,YWang2015}. In bulk form, this compound is a canonical superconducting member with a very simple structure~\cite{Wu08,Mao08}. It displays a typical tetragonal-to-orthorhombic structural transition, with $T_s \approx 90$ K, but, surprisingly, no N\'{e}el transition~\cite{McQueen2009,Medvedev2009,Bohmer2015,Baek2015,Nakayama14,Shimojima14,Watson15,Terashima15}. This is puzzling, because it differs from the standard case of the iron pnictides where the structural phase transition is accompanied by a $(\pi,0)$ antiferromagnetic (AFM) order~\cite{PCDai.2015}. Several theoretical proposals attribute this unusual behavior to the frustrated magnetism among the local moments \cite{YuSi_AFQ,FaWang2015,Glasbrenner2015}. Two of the present authors considered a generalized bilinear-biquadratic (GBQ) model on a square lattice and proposed that an antiferroquadrupolar (AFQ) state with wave vector $(\pi,0)$ describes the bulk FeSe~\cite{YuSi_AFQ}. This theoretical picture predicted low-energy spin excitations near $(\pi,0)$, which has since been experimentally observed~\cite{Rahn2015,WangZhao2016}. It also predicted a linear-in-energy spectral weight for such low-energy spin excitations and, over a wider energy range, spin excitations near both $(\pi,0)$ and $(\pi,\pi)$, all of which have also been verified in recent experiments~\cite{Wang2015,Shamoto2015}. More broadly, the neutron scattering measurements show that the spin spectral weight 
is even larger than that of the AFM state in the iron pnictides~\cite{Wang2015,Shamoto2015}, which provides further support for describing the magnetic properties of FeSe in terms of frustrated magnetism.

The proposed {\it two-sublattice} C$_4$-symmetry-breaking AFQ state is a novel state of matter, and systematic theoretical studies are clearly called for. Quadrupolar order {\it per se} in frustrated spin models has been studied before \cite{Blume1969, HHChen-quadrupole, Andreev_JETP, Papanicolaou1988, Shannon06,Lauchli2006,Toth2010,Toth2012, Bauer2012,Smerald2013, Smerald2015}, representing an intriguing spin state that involves the ordering of spin quadrupolar moments without exhibiting a magnetic dipolar order. However, two-sublattice AFQ order such as the proposed $(\pi,0)$ phase has not been realized before as a zero-temperature phase in such quantum-spin models, and the nature of the associated rotational symmetry breaking has not been addressed. In particular, it would be important to establish if the AFQ order is a true ground state of the GBQ model when the quantum fluctuations are fully accounted for. 

In this \textit{Letter}, we demonstrate that the $(\pi, 0)$ AFQ state is the ground state of the spin $S=1$ GBQ model on a square lattice over an extended parameter range. We have done so by two complementary means. We first show that the AFQ order has the lowest energy for a range of parameters based on a site-factorized wavefunction \cite{Lauchli2006,Toth2010,Bauer2012, Lai_SU(3)_gapless, Lai_SU(3)_chiral}. From a flavor-wave analysis, we show that quantum fluctuations lift an infinite degeneracy in the ground state energy
 and stabilize the AFQ ground state with order at either $(\pi,0)$ or $(0,\pi)$. 
Such order-from-disorder physics is analogous to what happens for the case of pure antiferromagnetic order~\cite{Henley1989, Chandra_Ising}, although it has never before been realized for any two-sublattice AFQ order.
We then show that the AFQ order is the true ground state even when the quantum fluctuations are treated fully and in an unbiased way, using the density matrix renormalization group (DMRG) method~ \cite{White-DMRG, White1993-DMRG}. 
Finally,  from a symmetry-based treatment, we establish that the AFQ order parameter does not couple to bilinear fermions, thereby demonstrating the consistency of the $(\pi,0)$ AFQ order with the single-electron spectrum observed in FeSe.
We stress that both the problem we address, and the analysis we carry through, are new to the present work. We note in passing that the stabilization of the C$_4$-symmetry-breaking AFQ by the quantum fluctuation effects not only provides an intriguing mechanism for the nematic order in the normal state of the iron chalcogenide FeSe, but also suggests the possible realization of such a ``hidden order" phase in cold atom systems tuned away from the SU(N$>$2) symmetric point, in which bilinear-biquadratic couplings can be realized~\cite{JYe_SU(N)_Science}.

\textit{Generalized Bilinear-Biquadratic Model}---
We consider the GBQ model on a two-dimensional square lattice,
\begin{eqnarray}\label{Eq:GBQmodel}
H= \sum_{i, \delta_n} \left[ J_n {\bf S}_i \cdot {\bf S}_j + K_n \left( {\bf S}_i \cdot {\bf S}_j \right)^2 \right],
\end{eqnarray}
where $j = i + \delta_n$, and $\delta_n$ connects site $i$ and its $n$th nearest neighbor sites with $n=1,~2,~3$. 
The couplings $J_n$ and $K_n$ are the bilinear and biquadratic couplings between the $n$th nearest neighbor spins. 
The importance of the biquadratic  couplings $K_n$ (along with the bilinear couplings $J_n$) has been suggested both from an analysis of the inelastic neutron-scattering spectra in the iron pnictides~\cite{Yu2012} as well as from {\it ab initio} studies~\cite{Wysocki2011}. The large magnitude inferred for the biquadratic coupling is compatible with the expectation for multi-orbital models in the bad-metal regime \cite{Fazekas1999}.
We expect that $K_n$ will contain not only a nearest-neighbor term ($n=1$) but also further-neighbor ones ($n>1$), in close analogy to the well-established case of $J_n$~\cite{Si2016}. A quadrupolar operator at site $i$, ${\bf Q}_i$, has five components: $Q^{x^2 - y^2}_i = (S^x_i)^2 - (S^y_i)^2$, $Q^{3z^2 - r^2}_i = [ 2 (S^z_i)^2 - (S^x_i)^2 - (S^y_i)^2]/\sqrt{3}$, $Q^{xy} = S^x_i S^y_i + S^y_i S^x_i$, $Q^{yz} =  S^y_iS^z_i +S^z_i S^y_i$, and $Q^{zx} = S^z_i S^x_i + S^x_i S^z_i$. The biquadratic term can be re-expressed as $({\bf S}_i \cdot {\bf S}_j )^2 = ({\bf Q}_i \cdot {\bf Q}_j) /2 - ({\bf S}_i \cdot {\bf S}_j) /2 + ({\bf S}^2_i {\bf S}^2_j)/3$. 

It is convenient to choose the time-reversal invariant basis of the SU(3) fundamental representation~\cite{Papanicolaou1988, Toth2012},
\begin{equation}\label{Eq:SU(3)_states}
\begin{aligned}
|x\ra = \frac{i |1\ra - i |\bar{1}\ra}{\sqrt{2}}, && |y\ra = \frac{ |1\ra + |\bar{1}\ra}{\sqrt{2}}, && |z\ra = -i |0\ra,
\end{aligned}
\end{equation}
where we abbreviate $|S^z = \pm1\ra \equiv |\pm1\ra$ $(|S^z = 0\ra \equiv |0\ra)$ and $|\bar{1}\ra \equiv |-1\ra$. We can introduce a site-factorized wavefunction at each site to characterize any ordered state with short-ranged correlations as
\begin{eqnarray}
|{\bf d}_i\ra = d^x_i |x\ra + d^y_i | y\ra + d^z_i |z\ra,
\end{eqnarray}
where $d^{x,y,z}_i$ are complex numbers and can be re-expressed in the vector form called director, ${\bf d}_i =  ( d^x_i, ~ d^y_i ,~ d^z_i  )$, with the basis $\{ |x\ra,~|y\ra,~|z\ra\}$. 
\begin{figure}[t]
   \centering
   \includegraphics[width=1.5 in]{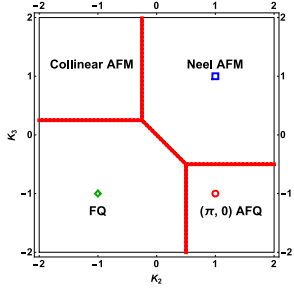}
   \caption{(Color online) The phase diagram derived from the site-factorized wavefunction studies, as a function of $K_2$ and $K_3$. We have fixed $J_1 = J_2 =1/4$, and $K_1 = -1$. Collinear AFM represents an antiferromagnet with wave vectors $(\pi,0)/(0,\pi)$, and N\'{e}el AFM with $(\pi,\pi)$. The $(\pi,0)$ AFQ corresponds to the antiferroquadrupolar phase, which has an infinite degeneracy that is to be lifted by quantum fluctuations. FQ corresponds to the ferroquadrupolar state. The red open circle, green open diamond, and blue open square are three parameter points for which the DMRG results will be presented.}
   \label{Fig:Phase_Diagram}
\end{figure}
We can then re-express the model Hamiltonian as \cite{supplemental}
\begin{eqnarray}\label{Eq:H_d}
H_{sf} = \sum_{i, \delta_n} \left[J_n\left| {\bf d}_i \cdot \bar{\bf d}_j\right|^2
+  \left(K_n - J_n\right) \left|{\bf d}_i \cdot {\bf d}_j \right|^2 + K_n\right],~~~~
\end{eqnarray}
where the subscript ``sf`` refers to the site-factorized Hamiltonian. In the following, we will drop the irrelevant constant terms in Eq.~\eqref{Eq:H_d}. Within the SU(3) basis, the ferroquadrupolar phase (FQ) has all directors aligned along a particular direction. In contrast, in AFQ the directors at different sublattices are orthogonal to each other.

\textit{AFQ order from site-factorized wavefunction}---
We study the phase diagram using a variational method based on the site-factorized wave-functions 
on a $L\times L$ square lattice with $L$ up to $6$
and periodic boundary condition. We first illustrate our result by considering fixed $J_1 = J_2 = 1/4$, $J_3=0$ and $K_1 = -1$, and variable $K_2$ and $K_3$. (See below about the robustness of our result over an extended parameter range.
As shown in Fig.~\ref{Fig:Phase_Diagram}, the ground state phase diagram contains four phases: a collinear AFM (CAFM) ordered at wave vectors $(\pi,0)/(0,\pi)$, a N\'{e}el AFM ordered at $(\pi,\pi)$, a FQ ordered at $(0,0)$, and an AFQ ordered at $(\pi,0)/(0,\pi)$. Within our approach, we did not find evidence for any three-sublattice AFQ order. \cite{Toth2010,Toth2012, Bauer2012}
\begin{figure}[t]
    \subfigure[]{\label{Fig:CollinearAFQ}\includegraphics[width= 1 in]{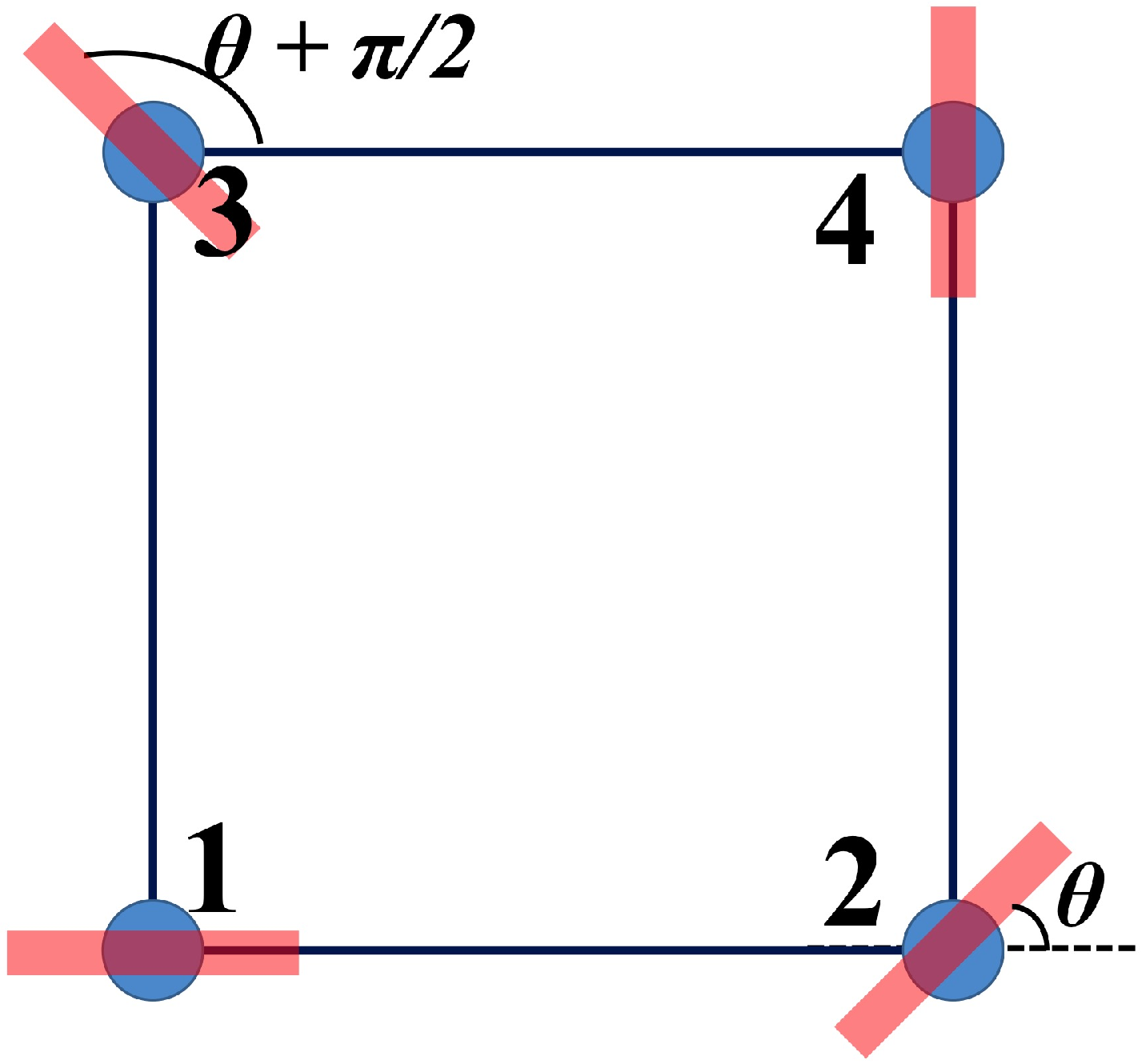}}
    \subfigure[]{\label{Fig:latticeAFQ}\includegraphics[width=1 in]{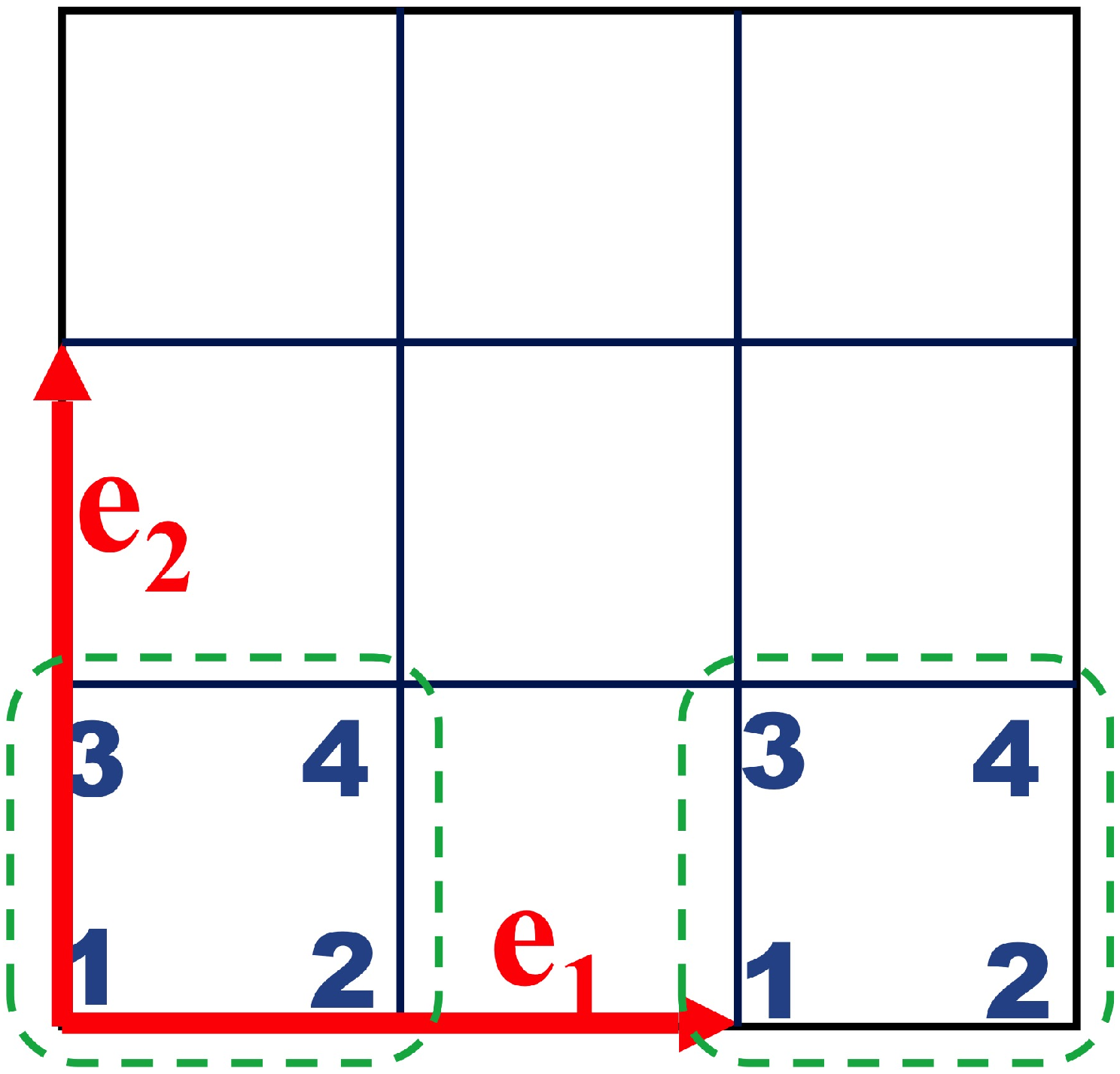}}
    \caption{(Color online) (a) Illustration of the $(\pi,0)$ AFQ found within the site-factorized wave function studies. The red bars are the directors ${\bf d}$ labeling the quadrupolar direction. Generically, there are 4 sublattices per unit cell, and the two independent directors are specified by one independent angle $\theta$. (b) Illustration of the square network consisting of the lattice for performing flavor-wave theory calculations. The unit cells, which contains $4$ sublattices, are connected by the vectors ${\bf e}_1 \equiv \hat{x} $ and ${\bf e}_2 \equiv \hat{y}$, where we set the lattice constant $a\equiv 1$.}
\label{Fig:CAFQ}
\end{figure}

Figure~\ref{Fig:CollinearAFQ} illustrates the directors in the $(\pi,0)$ AFQ, in which there are $4$ sublattices. The ${\bf d}$ directors connected by the second-neighbor bonds are \textit{mutually orthogonal} to each other, while the nearest-neighbor ${\bf d}$-s are subject to an angle $\theta$. In the $4$ sublattices, there are  only $2$ independent ${\bf d}$-s. We choose those sitting on sublattices $1$ and $2$ to be independent, which then specifies the ${\bf d}$-s on sites $3$ and $4$ straightforwardly due to orthogonality. This leads to the following parametrization for the ${\bf d}$-s:
\begin{equation}\label{Eq:parametrization_d}
\begin{array}{lr}
{\bf d}_1 = \begin{pmatrix} 1 &0 &0\end{pmatrix}, & {\bf d}_2= \begin{pmatrix} \cos\theta & \sin\theta & 0 \end{pmatrix},\\
{\bf d}_3 = \begin{pmatrix} -\sin\theta & \cos\theta& 0 \end{pmatrix}, & {\bf d}_4 = \begin{pmatrix}0 &1&0\end{pmatrix}.
\end{array}
\end{equation}
Despite the finite angle between ${\bf d}_1$ and ${\bf d}_2$, the energy of $(\pi,0)$ AFQ is \textit{independent} of the angle $\theta$ within this semiclassical approach, which can be seen by plugging ${\bf d}$ directors into Eq.~(\ref{Eq:H_d}). Thus, the semiclassical $(\pi,0)$ AFQ is \textit{infinitely degenerate} at the level of site-factorized wavefunction studies, which \textit{do not} include the quantum fluctuations. [Quantum fluctuations will lift the degeneracy (see below).] The boundaries between each phase can be determined analytically \cite{supplemental}, which are consistent with the numerical results.

\textit{Quantum fluctuations stabilizing $(\pi,0)$ AFQ}---
The $(\pi, 0)$ AFQ at the level of the site-factorized wave function is illustrated in Fig.~\ref{Fig:CollinearAFQ}, in which the angle $\theta$ varies between $0$ and $\pi$. The states with angles $\theta=0$ or $\pi/2$ correspond to the AFQ state of interest, at wave vector $(0,\pi)$ or $(\pi,0)$, respectively. Below we study the effect of the quantum fluctuations in this AFQ using the flavor-wave theory formulation.

For the flavor wave calculation, we associate $3$ Schwinger-bosons at each site $i$, $b_{i\alpha=x,y,z}$, to the states of Eq.~(\ref{Eq:SU(3)_states}), where $b^\dagger_{i\alpha} |vac\ra = |\alpha\ra$ with $|vac\ra$ being the vacuum state of the Schwinger bosons. The bosons satisfy a local constraint $\sum_{\alpha} b^\dagger_{i \alpha}b_{i \alpha} = 1$. The Hamiltonian, Eq.~(\ref{Eq:GBQmodel}), can be rewritten as
\begin{eqnarray}\label{Eq:Boson_H}
H = \sum_{i, \delta_n, \alpha, \beta} \left[ J_n b^\dagger_{i \alpha} b_{j \alpha} b^\dagger_{j\beta} b_{i\beta} + \left(K_n - J_n\right) b_{i\alpha}^\dagger b^\dagger_{j \alpha} b_{j\beta} b_{i\beta}\right].~~~~
\end{eqnarray}
Following the usual procedure of the spin-wave theory calculations, we introduce different local rotations around $z$-axis for each sublattice $i = 1,2,3,4$ as $a_{i \alpha} = \sum_{\beta} \left(\mathcal{R}^{\theta_i}_z\right)_{\alpha \beta} b_{i \beta}$, where $\mathcal{R}_z^{\theta_i}$ represents the SO(3) matrix for a rotation around the $z$-axis by angles $\theta_i$ that are determined according to Eq.~\eqref{Eq:parametrization_d} and Fig.~\ref{Fig:CollinearAFQ}. At each site, we assume that only $a_{ix}$ condenses, and we replace $a^\dagger_{ix}$ and $a_{ix}$ by $(M - a_{iy}^\dagger a_{iy} - a^\dagger_{iz} a_{iz})^{1/2}$, where $M=1$ in the present case. A $1/M$ expansion up to the quadratic order in the bosons $a_y$ and $a_z$ followed by an appropriate Holstein-Primakoff transformation allows us to extract the ground state energy. From now on we replace the labeling $a_{i\alpha} = a_{\alpha}({\bf r},a)$, where ${\bf r}$ runs over the Bravais lattice of unit cells of the square network and $a=1,2,3,4$ runs over the sub lattices, as illustrated in Fig.~\ref{Fig:latticeAFQ}. The different unit cells are connected by ${\bf e}_1 \equiv \hat{x}$ and ${\bf e}_2 \equiv \hat{y}$.

For clarity, we introduce $ D^T_\alpha({\bf k}) \equiv \{a_\alpha({\bf k},1), a_\alpha({\bf k},2), a_\alpha({\bf k},3), a_\alpha({\bf k},4)\}$ and $A^T_\alpha({\bf k}) \equiv \left\{ D^T_\alpha({\bf k}), D^\dagger_\alpha(-{\bf k})\right\}$, where $\alpha = y, z$. We arrange the Hamiltonian \cite{Xiao2009} to be $H = \mathcal{H}_c + \mathcal{H}_B$. The first term, $\mathcal{H}_c = 8 \sum_{k}\bigg{[}  J_1 +  K_1 +  J_2\left( 1 -\sin^2(2\theta)/8\right) -K_2+ J_3 \left(1-   \sum_{\mu=1,2} \cos({\bf k}\cdot {\bf e}_\mu ) + \sin^2(2\theta)/8\right)+ 3 K_3 \bigg{]}$, represents the semiclassical ground-state energy. The second term $\mathcal{H}_B$ is expressed as $\mathcal{H}_B = \sum_{{\bf k}, \eta = y,z} A_\eta^\dagger \mathcal{H}_{\eta} A_\eta $, with
\begin{eqnarray}\label{Eq:BosonH}
\mathcal{H}_{\eta} = \begin{pmatrix}
\alpha_{\eta} & \gamma_{\eta} \\
\gamma^\dagger_{\eta} & \alpha_{\eta}
\end{pmatrix},
\end{eqnarray}
where $\alpha_{\eta}$ and $\gamma_{\eta}$ are $4\times4$ Hermitian matrices and
are functions of momenta ${\bf k}$, couplings $J_{n}$ and $K_{n}$, and the angle $\theta$. $\mathcal{H}_B$ contains the zero-point energy of the boson fields, which plays the role of quantum correction to the semiclassical ground-state energy. We leave the full expressions of the matrices to the Supplemental Material \cite{supplemental}.
 
\begin{figure}[t]
   \centering
   \includegraphics[width=1.5 in ]{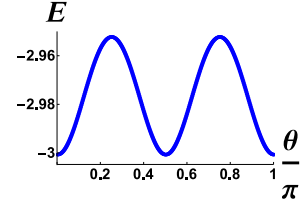}
   \caption{(Color online) Energy per site of the $(\pi,0)$ AFQ vs $\theta$ obtained in the flavor-wave theory calculations. The system we use in the numerics consists of 100$\times$100 unit cells, with $4$ sites per unit cell. The parameters for this calculation are $(K_2, K_3) = (1,-1)$.
}
   \label{Fig:FLWT_AFQ}
\end{figure}

Figure~\ref{Fig:FLWT_AFQ} shows the ground state energy of the $(\pi, 0)$ AFQ vs $\theta$ within the flavor-wave theory at $\{K_2, K_3\} = \{1, -1\}$. The two degenerate quadrupolar ground states at $\theta= 0, \pi/2$ correspond to the AFQ with the ordering wavevector $(\pi,0)$ or $(0,\pi)$. We conclude that the quantum fluctuations lift the infinite degeneracy and stabilize the $(\pi,0)$ AFQ.

\textit{Density Matrix Renormalization Group Analysis}---
To further demonstrate the stability of the $(\pi,0)$ AFQ phase and to analyze the GBQ model in an unbiased way, we turn next to the study of the ground states using the SU(2) DMRG calculations~\cite{White-DMRG,mcculloch2002,gong2014square,gong2014kagome, GongJ1J213}. 
To search for the $(\pi,0)$ AFQ order, we specifically consider the parameter point, $(K_2,K_3)=(1,-1)$, where the $(\pi,0)$ AFQ is realized in Fig.~\ref{Fig:Phase_Diagram} (recall $J_1=J_2=1/4$ and $K_1=-1$). For comparison, we also consider two parameter points in the nearby regimes, $(K_2,K_3) = (-1,-1)$ and $(1,1)$, corresponding to the FQ and N\'{e}el AFM, respectively, in Fig.~\ref{Fig:Phase_Diagram} (the other parameters are unchanged). We perform DMRG simulations on cylindrical geometries with $L_y=6,8$ lattice spacings keeping up to $6000$ SU(2) states and $L_y=10$ keeping up to $4000$ SU(2) states. We rescale the parameters with respect to $J_1$. The largest truncation errors are around $10^{-5}$. Especially on the $L_y=10$ cylinder, we have checked the results corresponding to $2000$ and $4000$ SU(2) states, and found the differences to be small (around $10^{-3}$) for $m^2_S$ and $m^2_Q$.

\begin{figure}[t]
    {\includegraphics[width= 3 in]{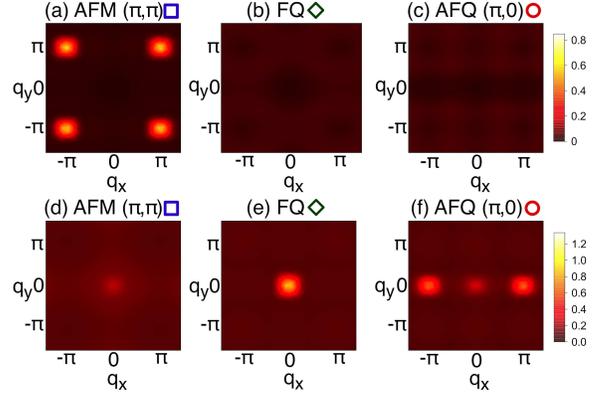}}
    \caption{(Color online) The spin dipolar ((a)-(c)) and quadrupolar ((d)-(f)) structure factors, $m_s^2({\bf q})$ and $m_Q^2({\bf q})$, with $(J_1,J_2,J_3,K_1)=(1/4, 1/4, 0, -1)$ on $L_y=8$ lattice for N\'{e}el order [(a)/(d)] with $(K_2,K_3) = (1,1)$ ), FQ [(b)/(e)] with $(K_2,K_3) = (-1,-1$), and $(\pi,0)$ AFQ [(c)/(f)] with $(K_2,K_3) = (1,-1)$. The color represents the peak height of the corresponding structure factor. 
    }
 \label{Fig:DMRG}
\end{figure}
We choose $L_y = L$ to calculate the spin ($\left\la {\bf S}_i \cdot {\bf S}_j\right\ra$)
and quadrupolar ($\left\la {\bf Q}_i \cdot {\bf Q}_j \right\ra$) correlation functions
in the middle of $L\times 2L$ cylinder systems to obtain the corresponding
structure factors \cite{gong2014square, GongJ1J213}, $m^{2}_{S}({\bf q})\equiv
(1/L^4)\sum_{ij}\la {\bf S}_{i}\cdot {\bf S}_{j}\ra e^{i{\bf q}\cdot({\bf r}_i-{\bf r}_j)}$
and $m^{2}_{Q}({\bf q})\equiv (1/L^4)\sum_{ij}\la {\bf Q}_{i}\cdot {\bf Q}_{j}\ra
e^{i{\bf q}\cdot({\bf r}_i - {\bf r}_j)}$, in Figs.~\ref{Fig:DMRG}(a)-(f). 
We obtain the results at parameter points $(K_2, K_3) = (1,1),~(-1,-1),$ and $(1,-1)$ shown, respectively, in Figs.~\ref{Fig:DMRG}[(a)/(d)], [(b)/(e)], and [(c)/(f)]. Fig.~\ref{Fig:DMRG}[(a)/(d)] show a sharp peak at $(\pm\pi,\pm\pi)$ in $m^2_S({\bf q})$ and a weak FQ peak at $(0,0)$ in $m^2_{Q}({\bf q})$ suggesting the N\'{e}el AFM. We note that for spin-$1$ system the magnetically-ordered states are expected to show finite FQ order. Fig.~\ref{Fig:DMRG}[(b)/(e)] show no magnetic order signature in $m^2_S({\bf q})$ and a sharp peak at $(0,0)$ in $m^2_Q({\bf q})$ suggesting the ground state is FQ.  Fig.~\ref{Fig:DMRG}[(c)/(f)] show no clear signature in $m^2_S({\bf q})$ and sharp peaks at $(\pm\pi,0)$ in $m^2_{Q}({\bf q})$ suggesting the realization of the $(\pi,0)$ AFQ, which is confirmed under finite-size scaling analysis. \cite{supplemental} We note that Fig.~\ref{Fig:DMRG}(f) also shows a peak at ${\bf q} = (0,0)$ in $(\pi,0)$ AFQ. This is theoretically expected: For a two-sublattice AFQ order at $(\pi,0)$, one diagonal component of the quadrupolar operator $Q^{x^2-y^2}$ takes staggered values at sublattices $A$ and $B$, $\la Q_i^{x^2 -y^2}\ra =\la (S_i^x)^2\ra -\la (S_i^y)^2\ra = (-1)^i$, which implies $\la (S_i^x)^2 \ra = \delta_{i A}$  and $\la(S_i^y)^2\ra  = \delta_{i B}$ ; correspondingly, the other diagonal component takes \textit{uniform} expectation values at each site, $\la Q^{3z^2 - r^2}_i \ra = -1/\sqrt{3}$, and thus shows the FQ peak.

\textit{Discussions}---
We close by remarking on several points. First, both our analytical and numerical calculations indicate that the $(\pi,0)$ AFQ order is not accompanied by any AFM order. 

Second, the $(\pi,0)$ AFQ ground state is stable over a very wide range in the parameter space. To illustrate this point, we consider the case of $-K_1/J_2=0.8$, which is expected to be realistic to FeSe since it is already close to that extracted from fitting the spin spectra of related iron-based systems~\cite{Yu2012}. Continuing to set $J_1=J_2=1/4$, and taking $K_2 = -K_3 = -K_1 =1/5$, we show that the $(\pi,0)$ AFQ ground state persists (see the Supplemental Material; particularly, Fig.~S3) \cite{supplemental}.

Third, the $(\pi,0)$ AFQ state breaks the C$_4$ symmetry, and associated with it is an Ising-nematic order. The latter is expected to be dominated by the following order parameter~\cite{YuSi_AFQ}:
\begin{equation}
 \sigma_2 = \sum_i \left [ (\mathbf{S}_i\cdot\mathbf{S}_{i+\hat{x}})^2
 - (\mathbf{S}_i\cdot\mathbf{S}_{i+\hat{y}})^2 \right ].
\end{equation}
While this is clearly the case for the ground state, $\sigma_2$ will persist at nonzero temperatures even in the purely two-dimensional limit. (In the presence of an interlayer coupling, the AFQ order will also extent to nonzero temperatures.) This provides the basis to understand the nematic transition at $T_s$ in FeSe.

Fourth, in a $(\pi,0)$ AFQ state, the low-energy spin excitations are expected to be 
concentrated near the wavevector $(\pi,0)$. The spectral weight at low energies 
should be linear in $\omega$ ~\cite{YuSi_AFQ}:
It is proportional to $[M(\omega)]^2/ \omega$, 
with the spectral weight of the quadrupolar Goldstone mode {\it per se} contributing the factor $1/\omega$, and the spin dipolar matrix element of the quadrupolar mode $M(\omega)$ being $ \propto \omega$. (This argument is valid for any AFQ order at zero magnetic field and, indeed, the linear-in-$\omega$ dependence also appears in the three-sublattice $(2\pi/3, 2\pi/3)$ AFQ state on the triangular lattice  \cite{Smerald2013}.) Such a linear dependence has been observed (up to about $50$ meV) by the recent neutron-scattering experiments in FeSe\cite{Wang2015,Shamoto2015}. At higher energies, the spin excitations are expected to spread over a large range of wavevectors, including a sizable spectral weight near ($\pi$,$\pi$). This is also consistent with the neutron-scattering measurements in FeSe \cite{Wang2015,Shamoto2015}.
 
Finally, the quadrupolar operator acts like a spin-$2$ operator. Thus, in the absence of spin-orbit coupling, the AFQ order parameter cannot be coupled to the bilinear fermion fields. (In the Supplemental Material \cite{supplemental}, the result is derived from a rigorous group-symmetry analysis.) This implies that the AFQ order does not reconstruct the Fermi surface. Instead, the coupling to the bilinears of the itinerant electrons is only through the nematic order parameter, which induces a distortion of the Fermi surface. In contrast to what happens above the ordering temperature, the Fermi surface in the AFQ state will lose the invariance under a C$_4$-rotation: {\it e.g.}, the hole Fermi pockets near $\Gamma$ will be elongated along one of the axis directions. All these features are consistent with the observations of photoemission experiments\cite{Watson15}, when twin domains are taken into account.

\textit{Acknowledgement}--
We would like to acknowledge useful discussions with Shoushu Gong and Zhentao Wang. We thank Zhentao Wang for providing the codes for site-factorized wavefunction studies and Shoushu Gong for providing the SU(2) DMRG code. This work was supported in part by the NSF Grant No.\ 
DMR-1611392 and the Robert A.\ Welch Foundation Grant No.\ C-1411 (H-H.L., W-J.H. and Q.S.), by the NSF Grant No.\ DMR-1350237 (H-H. L. and W-J.H.), by a Smalley Postdoctoral Fellowship in Rice Center for Quantum Materials (H-H. L.), and by the National Science Foundation of China Grant number 11374361 and the Fundamental Research Funds for the Central Universities and the Research Funds of Renmin University of China (R.Y.). The majority of the computational calculations have been performed on the Shared University Grid at Rice funded by NSF under Grant EIA-0216467, a partnership between Rice University, Sun Microsystems, and Sigma Solutions, Inc., the Big-Data Private-Cloud Research Cyber-infrastructure MRI-award funded by NSF under Grant No. CNS-1338099, the Extreme Science and Engineering Discovery Environment (XSEDE) by NSF under Grants No. DMR160003.

\bibliography{biblio4AFQ}

\end{document}


\title{Supplemental Material for Antiferroquadrupolar order and rotational symmetry breaking in a generalized bilinear-biquadratic model on a square lattice} 
\author{Hsin-Hua Lai}
\affiliation{Department of Physics and Astronomy \& Rice Center for Quantum Materials, Rice University, Houston, Texas 77005, USA}
\author{Wen-Jun Hu}
\affiliation{Department of Physics and Astronomy \& Rice Center for Quantum Materials, Rice University, Houston, Texas 77005, USA}
\author{Emilian M. Nica}
\affiliation{Department of Physics and Astronomy \& Rice Center for Quantum Materials, Rice University, Houston, Texas 77005, USA}
\author{Rong Yu}
\affiliation{Department of Physics, Renmin University of China, Beijing, 100872, China}
\affiliation{Department of Physics and Astronomy, Shanghai Jiao Tong
University, Shanghai 200240, China and Collaborative Innovation
Center of Advanced Microstructures, Nanjing 210093, China}
\author{Qimiao Si}
\affiliation{Department of Physics and Astronomy \& Rice Center for Quantum Materials, Rice University, Houston, Texas 77005, USA}
\date{\today}
\begin{abstract}
In this supplemental material, we give more details of the derivations, the real-space data and the finite-size scaling of the density matrix renormalization group in $(\pi,0)$ antiferroquadrupolar phase, and symmetry group analysis of the couplings between local moments and multi-orbital conduction electrons in the absence of spin-orbit couplings.
\end{abstract}

\maketitle
\section{Phase boundaries within site-factorized wavefunction studies}
For the site-factorized wavefunction studies, it is convenient to choose the time-reversal invariant basis of the SU(3) fundamental representation \cite{Papanicolaou1988, Toth2012},
\begin{equation}
\begin{aligned}
|x\ra = \frac{i |1\ra - i |\bar{1}\ra}{\sqrt{2}}, && |y\ra = \frac{ |1\ra + |\bar{1}\ra}{\sqrt{2}}, && |z\ra = -i |0\ra,
\end{aligned}
\end{equation}
where we abbreviate $|S^z = \pm1\ra \equiv |\pm1\ra$ $(|S^z = 0\ra \equiv |0\ra)$ and $|\bar{1}\ra \equiv |-1\ra$. The site-factorized wavefunctions at each site
can be introduced to characterize any possible ordered state with short-ranged correlations as
\begin{eqnarray}
|{\bf d}_i\ra = d^x_i |x\ra + d^y_i | y\ra + d^z_i |z\ra,
\end{eqnarray}
where $d^x_i$, $d^y_i$, $d^z_i$ are complex numbers and can be re-expressed in the vector form with the basis $\{ |x\ra,~|y\ra,~|z\ra\}$ as $ {\bf d}_i =  ( d^x_i ~ d^y_i ~ d^z_i  )$. It is convenient to separate the real and imaginary parts of ${\bf d}_j$ as${\bf d}_i = {\bf u}_i + i {\bf v}_i$. The normalization of the wavefunction leads to the constraint ${\bf d}_i \cdot \bar{\bf d}_i = 1$, or equivalently, ${\bf u}^2_i + {\bf v}^2_i =1$, and the overall phase can be fixed by requiring ${\bf d}_i^2 = \bar{\bf d}_i^2$, i.e., ${\bf u}_i \cdot {\bf v}_i = 0$. In a pure quadrupolar state, ${\bf d}$ will take either a real or imaginary value, but not both, and the associated director is parallel to the director vector ${\bf d}$. This is to be contrasted with a magnetic order, for which ${\bf d}$ contains both real and imaginary components. Within the spin-coherent state framework, we can determine the spin operator from ${\bf S}_i = 2 {\bf u}_i \times {\bf v}_i$. In terms of the components of the ${\bf d}$, the spin and quadrupole operators can be written as $S^\alpha = - i \sum_{\beta \gamma} \epsilon^{\alpha \beta \gamma} \bar{d}^\beta d^\gamma$,  $Q^{x^2 - y^2} = - |d^x|^2 + |d^y|^2$, $Q^{3z^2 - r^2} = \left[ |d^x|^2 + |d^y|^2 -2 |d^z|^2 \right]/\sqrt{3}$, $Q^{\alpha \beta}|_{\alpha \not=\beta} = - \bar{d}^\alpha d^\beta - \bar{d}^\beta d^\alpha$. We can then re-express the bilinear-biquadratic model Hamiltonian that we focus on in this \textit{Letter} as 
\begin{eqnarray} \label{Eq:SF_H}
H = \sum_{i, \delta_n} \left[J_n\left| {\bf d}_i \cdot \bar{\bf d}_j\right|^2
+  \left(K_n - J_n\right) \left|{\bf d}_i \cdot {\bf d}_j \right|^2 + K_n\right],
\end{eqnarray}
where $j = i + \delta_n$, and $\delta_n$ connects site $i$ and its $n$th nearest neighbor sites with $n=1,~2,~3$. 

Within the site-factorized wavefunction studies, the phase boundaries between each phase can be determined analytically based on Eq.~\eqref{Eq:SF_H}. Focusing on the energy per site, we find that the site energies for N\'{e}el AFM, collinear AFM (CAFM), FQ, and $(\pi,0)$ AFQ (ignore the constant terms) are
\begin{eqnarray}
&& \mathcal{E}^{N\acute{e}el}_{AFM} = 2 \left(K_1 - J_1\right) + 2 J_2 + 2 J_3,\\
&& \mathcal{E}_{CAFM} = K_1 + 2 (K_2 - J_2) + 2J_3,\\
&& \mathcal{E}_{FQ} = 2\left( K_1 + K_2 + K_3 \right),\\
&& \mathcal{E}_{(\pi,0)~AFQ} = K_1 + 2K_3.
\end{eqnarray}
We can then determine the boundaries between each phase based on these energies
\begin{enumerate}
\item[(1)]: Phase boundary between N\`{e}el AFM and CAFM is
\begin{eqnarray}
K_2^{N\acute{e}el~AFM/CAFM} = \frac{1}{2}\left( K_1 - 2 J_1 + 4 J_2\right).
\end{eqnarray}
For the parameters used in the main texts, the boundary is at $K_2^{N\acute{e}el~AFM/CAFM} = -1/4$.
\item[(2)]: Phase boundary between FQ and $(\pi,0)$ AFQ is
\begin{eqnarray}
K_2^{FQ/(\pi,0)~AFQ} = -\frac{1}{2} K_1,
\end{eqnarray}
which corresponds to $K_2^{FQ/(\pi,0)~AFQ} = 1/2$ in the parameters used in main texts.\item[(3)]: Phase boundary between FQ and CAFM is
\begin{eqnarray}
K_3^{FQ/CAFM} = \frac{1}{2}\left( -K_1 - 2J_2 + 2J_3\right),
\end{eqnarray}
which is $K_3^{FQ/CAFM} = 1/4$  in the parameters used in main texts.
\item[(4)]: Phase boundary between $(\pi,0)$ AFQ and N\`{e}el AFM is
\begin{eqnarray}
K_3^{(\pi,0)AFQ/N\acute{e}el~AFM} = \frac{1}{2}K_1 - J_1 + J_2 + J_3,
\end{eqnarray}
which is $K_3^{(\pi,0)AFQ/N\acute{e}el~AFM} = -1/2$ in the parameters used in main texts.
\item[(5)]: Phase boundary between FQ and N\'{e}el AFM can be determined by the equation
\begin{eqnarray}
&& K^{FQ/N\acute{e}el~AFM}_2 + K^{FQ/N\acute{e}el~AFM}_3 = J_2 + J_3 - J_1,
\end{eqnarray}
which is $K^{FQ/N\acute{e}el~AFM}_2 + K^{FQ/N\acute{e}el~AFM}_3 = 0$ in the parameters used in main texts.
\end{enumerate}
Before closing the discussions of the site-factorized wavefunction analysis in this \textit{Letter}, we remark that a finite $K_3$ is important for the stability of the $(\pi,0)$ AFQ. At the level of site-factorized wavefunction studies, without $K_3$ we \textit{do not} find any parameter regime of $(\pi,0)$ AFQ. In addition, The $(\pi,0)$ AFQ is purely quadrupolar at each lattice site. This is to be contrasted with the semiordered phase discussed in Refs.~\cite{Papanicolaou1988,Toth2012}, in which 
one sublattice is purely quadrupolar while the other can be purely magnetic, purely quadrupolar or of mixed character. For a further check, we perform the site-factorized wavefunction analysis up to a $6\times6$ lattice inside the $(\pi,0)$ AFQ regime and find a consistent result.
\clearpage
\section{Flavor wave theory for $(\pi,0)$ AFQ}
Within the flavor wave calculation \cite{Papanicolaou1988, Toth2012} for the $(\pi,0)$ AFQ with $4$ sublattices per unit cell, we  associate $3$ Schwinger-bosons at each site $i$, $b_{i\alpha=x,y,z}$, to the states under SU(3) time-reversal invariant basis, $|x\ra,~|y\ra,~|z\ra$, where $b^\dagger_{i\alpha} |vac\ra = |\alpha\ra$ with $|vac\ra$ being the vacuum state of the Schwinger bosons. The bosons satisfy a local constraint
\begin{eqnarray}\label{Eq:constraint}
\sum_{\alpha} b^\dagger_{i \alpha}b_{i \alpha} = 1
\end{eqnarray}
The spin and quadrupole operators can be written bilaterally in the Schwinger bosons,
\begin{eqnarray}
&& S^\alpha_i = -i\epsilon_{\alpha \beta \gamma} b_{i \beta}^\dagger b_{i \gamma},\\
&& Q^{x^2 - y^2}_i = -b^\dagger_{ix}b_{ix} +b^\dagger_{iy}b_{iy}, \\
&& Q^{3z^2 - r^2}_i = (b_{ix}^\dagger b_{ix} + b_{iy}^\dagger b_{iy} - 2 b^\dagger_{iz}b_{iz})/\sqrt{3}, \\
&& Q^{xy}_i = - b^\dagger_{ix}b_{iy} - b^\dagger_{i y}b_{i x},\\
&& Q^{yz}_i = - b^\dagger_{iy}b_{iz} - b^\dagger_{i z}b_{i y},\\
&& Q^{zx}_i = - b^\dagger_{iz}b_{ix} - b^\dagger_{i x}b_{i z}.
\end{eqnarray}
The GBQ model Hamiltonian in terms of spin operators can be re-expressed in terms of the bosons,
\begin{eqnarray}
H = \sum_{i, \delta_n} \left[ J_n {\bf S}_i \cdot {\bf S}_j + K_n \left( {\bf S}_i \cdot {\bf S}_j \right)^2 \right] = \sum_{i, \delta_n, \alpha, \beta} \left[ J_nb^\dagger_{i \alpha} b_{i \alpha} b^\dagger_{j\beta} b_{i\beta} + \left(K_n - J_n\right) b_{i\alpha}^\dagger b^\dagger_{j \alpha} b_{i\beta} b_{i\beta}\right],
\end{eqnarray}
where $j = i + \delta_n,$ and $\delta_n$ (with $n = 1,~2,~3$) connects site $i$ to its $n$th nearest neighbor sites. For performing flavor wave theory calculation, we introduce different local rotations for site $j$ as
\begin{eqnarray}
\begin{pmatrix}
a_{ix} \\
a_{iy} \\
a_{iz}
\end{pmatrix} =
\begin{pmatrix}
\cos\theta_i & \sin\theta_i & 0 \\
-\sin\theta_i & \cos\theta_i & 0\\
0 & 0 & 1
\end{pmatrix}
\begin{pmatrix}
b_{ix}\\
b_{iy}\\
b_{iz}
\end{pmatrix},
\end{eqnarray}
which still preserves the local constraint of Eq.~(\ref{Eq:constraint}) with $b_{i\alpha} \rightarrow a_{i\alpha}$. At each site only one flavor of bosons $a_{ix}$ condenses, and we replace $a^\dagger_{ix}$ and $a_{ix}$ by $(M - a_{iy}^\dagger a_{iy} - a^\dagger_{iz} a_{iz})^{1/2}$, with $M=1$ in the present case. A $1/M$ expansion up to the quadratic order of the  bosons $a_y$ and $a_z$ followed by an appropriate transformation allows us to extract the ground state energy. Within the $(\pi,0)$ AFQ, we fix the angles $\theta_1 = 0$, $\theta_4=\pi/2$, and make the angle site 2 general, $\theta_2 = \theta$, and $\theta_3 = \theta  + \pi/2$. We will replace the labeling $a_{i\alpha} = a_{\alpha}({\bf r},a)$, where ${\bf r}$ runs over the Bravais lattice of unit cells of the square network and $a=1,2,3,4$ runs over the sub lattices. The different unit cells are connected by ${\bf e}_1 \equiv \hat{x}$ and ${\bf e}_2 \equiv \hat{y}$. In the $d$ boson language, we have
\begin{equation}
\begin{array}{lr}
\Bigg{\{}
\begin{array}{l}
b_x(\vec{r},1) = a_x(\vec{r},1), \\
b_y(\vec{r},1),=a_y(\vec{r},1), \\
b_z(\vec{r},1) = a_z(\vec{r},1);
\end{array} & \hspace{1cm}
\Bigg{\{}
\begin{array}{l}
b_x(\vec{r},2) = \cos(\theta) a_x(\vec{r},2) - \sin(\theta) a_y(\vec{r},2),\\
b_y(\vec{r},2) = \sin(\theta) a_x(\vec{r},2) + \cos(\theta) a_y(\vec{r},2), \\
b_z(\vec{r},2) = a_z(\vec{r},2);
\end{array}
\end{array}
\end{equation}
\begin{equation}
\begin{array}{lr}
\Bigg{\{}
\begin{array}{l}
b_x(\vec{r},3) = - \sin(\theta) a_x(\vec{r},3) - \cos(\theta) a_y(\vec{r},3),\\
b_y(\vec{r},3) = \cos(\theta) a_x(\vec{r},3) - \sin(\theta) a_y(\vec{r},3),\\
b_z(\vec{r},3) = a_z(\vec{r},3);
\end{array} & \hspace{0.6 cm} 
\Bigg{\{}
\begin{array}{l}
b_x(\vec{r},4) = - a_y(\vec{r},4),\\
b_y(\vec{r},4) = a_x(\vec{r},4),\\
b_z(\vec{r},4) = a_z(\vec{r},4).
\end{array}
\end{array}
\end{equation}

We below introduce $ D^T_{\alpha = y,z} ({\bf k}) = \{a_\alpha({\bf k},1), a_\alpha({\bf k},2), a_\alpha({\bf k},3), a_\alpha({\bf k},4)\}$, and $A_\alpha({\bf k}) = \left\{ D^T_\alpha({\bf k}), D^\dagger_\alpha(-{\bf k})\right\}$. Within the linear flavor wave theory calculation, we find that the Hamiltonian is $H = H_c + H_B$, where
\begin{eqnarray}
H_c = 8 \sum_{k}\bigg{[}  J_1 +  K_1 +  J_2\left( 1 -\frac{\sin^2(2\theta)}{8}\right) -  K_2 +  J_3 \left(1-   \sum_{\mu} \cos({\bf k}\cdot {\bf e}_\mu ) + \frac{\sin^2(2\theta)}{8}\right)+ 3 K_3 \bigg{]},
\end{eqnarray}
and can be determined straightforwardly independent of the boson fields. The $H_B \equiv \sum_{n=1,2,3}H_n$ represent the boson Hamiltonian related to the $n$th neighbor couplings. After Fourier transform, we find that $H_n = \sum_{{\bf k}, \eta = y,z} A_\eta^\dagger \mathcal{H}_{n,\eta} A_\eta $, with
\begin{eqnarray}
\mathcal{H}_{n,\eta} = \begin{pmatrix}
\alpha_{n,\eta} & \gamma_{n,\eta} \\
\gamma^\dagger_{n,\eta} & \alpha_{n,\eta}
\end{pmatrix},
\end{eqnarray}
where $\alpha_{n,\eta}$ and $\gamma_{n,\eta}$ are $4\times4$ Hermitian matrices which are detailed below.\\

\textit{Nearest-neighbor terms}--
 After Fourier transform, we find that the nearest-neighbor terms in the momentum space as $H_1 = \sum_{{\bf k},\eta}  A_\eta^\dagger \mathcal{H}_{1,\eta} A_\eta$, with
\begin{eqnarray}
&&\alpha^{11}_{1,y}=\alpha^{22}_{1,y} = \alpha^{33}_{1,y} = \alpha^{44}_{1,y}=\alpha^{14}_{1,y}=\alpha^{23}_{1,y}=0, \\
&& \alpha^{12}_{1,y} = \alpha^{34}_{1,y}=\frac{1}{2}\left( J_1 - K_1 \sin^2\theta \right)\left(1 + e^{-i {\bf k}\cdot {\bf e}_1}\right), \\
&& \alpha^{13}_{1,y} = \alpha^{24}_{1,y} =\frac{1}{2}\left( J_1 - K_1 \cos^2\theta \right)\left( 1+ e^{-i {\bf k}\cdot {\bf e}_2}\right),
\end{eqnarray}
with $\alpha^{ab}_{1,y} = \left(\alpha^{ba}_{1,y}\right)^*$, and
\begin{eqnarray}
&& \gamma_{1,y}^{11} = \gamma_{1,y}^{22} = \gamma_{1,y}^{33} = \gamma_{1,y}^{44} = \gamma_{1,y}^{14} = \gamma_{1,y}^{23}=0,\\
&& \gamma_{1,y}^{12} = \gamma_{1,y}^{34} = \frac{1}{2}\left( K_1 \cos^2\theta -J_1\right)\left( 1+ e^{-i {\bf k}\cdot {\bf e}_1}\right),\\
&& \gamma_{1,y}^{13} = \gamma_{1,y}^{24} = \frac{1}{2}\left( K_1 \sin^2\theta - J_1 \right) \left( 1+ e^{-i {\bf k}\cdot {\bf e}_2}\right),
\end{eqnarray}
with $\gamma^{ab}_{1,y} = \left( \gamma^{ba}_{1,y}\right)^*$. The matrix $\mathcal{H}_{1,z}$ can be characterized by the elements
\begin{eqnarray}
&& \alpha_{1,z}^{11} = \alpha_{1,z}^{22} = \alpha_{1,z}^{33} = \alpha_{1,z}^{44} = -K_1,\\
&& \alpha_{1,z}^{12}= \alpha_{1,z}^{34} = \frac{J_1\cos\theta}{2}\left( 1 + e^{-i {\bf k}\cdot {\bf e}_1}\right),\\
&& \alpha_{1,z}^{13} =-\alpha_{1,z}^{24}= - \frac{J_1 \sin\theta}{2}\left( 1 + e^{-i {\bf k}\cdot {\bf e}_2}\right),\\
&& \alpha_{1,z}^{14} = \alpha_{1,z}^{23} = 0,
\end{eqnarray}
and
\begin{eqnarray}
&& \gamma_{1,z}^{11} = \gamma_{1,z}^{22} = \gamma_{1,z}^{33} = \gamma_{1,z}^{44} = \gamma_{1,z}^{14} = \gamma_{1,z}^{23} = 0,\\
&& \gamma_{1,z}^{12} = \gamma_{1,z}^{34}=\frac{K_1 - J_1}{2}\cos\theta \left[ 1 + e^{-i {\bf k}\cdot {\bf e}_1}\right],\\
&& \gamma_{1,z}^{13} = - \gamma^{24}_{1,z} = \frac{J_1 - K_1}{2}\sin\theta \left[ 1 + e^{-i {\bf k}\cdot {\bf e}_2} \right],
\end{eqnarray}
for fully characterizing $\mathcal{H}_{1,z}$.\\

\textit{Second-neighbor terms}--
After Fourier transform, we find that $H_2 = \sum_{{\bf k},\eta}  A_\eta^\dagger \mathcal{H}_{2,\eta} A_\eta$, with
\begin{eqnarray}
&& \alpha_{2,y}^{11} = \alpha_{2,y}^{44} = 2K_2,\\
&& \alpha_{2,y}^{22} = \alpha_{2,y}^{33} = 2K_2+ \frac{J_2}{2}\sin^2(2\theta),\\
&& \alpha_{2,y}^{14} = \frac{ J_2 - K_2}{2}  \left[ 1 + e^{-i {\bf k}\cdot{\bf e}_1} + e^{-i {\bf k}\cdot {\bf e}_2} + e^{-i {\bf k}\cdot({\bf e}_1 + {\bf e}_2)}\right],\\
&& \alpha_{2,y}^{23} = \frac{J_2 - K_2 }{2} \left[ 1 + e^{i {\bf k}\cdot {\bf e}_1} + e^{-i {\bf k}\cdot {\bf e}_2} + e^{i {\bf k}\cdot ({\bf e}_1 - {\bf e}_2)}\right],\\
&& \alpha_{2,y}^{12} = \alpha_{2,y}^{13}=\alpha_{2,y}^{24} = \alpha_{2,y}^{34} = 0,
\end{eqnarray}
and
\begin{eqnarray}
\nonumber && \gamma_{2,y}^{11} = \gamma_{2,y}^{22} = \gamma_{2,y}^{33} = \gamma_{2,y}^{44} = \gamma_{2,y}^{12}=\gamma_{2,y}^{13}=\gamma_{2,y}^{24} =\gamma_{2,y}^{34}=0,\\ \\
&& \gamma_{2,y}^{14} = -\frac{J_2}{2} \left( 1 + e^{-i {\bf k}\cdot {\bf e}_1} + e^{-i {\bf k}\cdot {\bf e}_2} + e^{-i {\bf k} \cdot ({\bf e}_1 + {\bf e}_2)}\right),\\
&& \gamma_{2,y}^{23} = -\frac{J_2}{2}\left( 1 + e^{i {\bf k}\cdot {\bf e}_1} + e^{-i {\bf k}\cdot {\bf e}_2} + e^{i {\bf k}\cdot({\bf e}_1 - {\bf e}_2)}\right),
\end{eqnarray}
for fully characterizing the $\mathcal{H}_{2,y}$. Similarly we have obtained
\begin{eqnarray}
\mathcal{H}_{2,z} = 0,
\end{eqnarray}
We note that $\mathcal{H}_{2,z}=0$ is due to the choice we made for charactering the $(\pi,0)$ AFQ.\\

\textit{Third neighbor terms}--
After Fourier transform, we find that $H_3 = \sum_{{\bf k},\eta} A_\eta^\dagger \mathcal{H}_{3,\eta} A_\eta$, with
\begin{eqnarray}
&& \alpha_{3,y}^{11} = \alpha_{3,y}^{44} = -2K_3, \\
&& \alpha_{3,y}^{22} = \alpha_{3,y}^{33} = J_3\sum_\mu \cos({\bf k}\cdot {\bf e}_\mu) - \frac{J_3}{2}\sin^2(2\theta) - 2K_3,\\
&& \alpha_{3,y}^{12}= \alpha_{3,y}^{13} = \alpha_{3,y}^{14} = \alpha_{3,y}^{23} =\alpha_{3,y}^{24} = \alpha_{3,y}^{34} =0,
\end{eqnarray}
and
\begin{eqnarray}
&& \gamma_{3,y}^{11} = \gamma_{3,y}^{22} = \gamma_{3,y}^{33} = \gamma_{3,y}^{44} = \left( K_3 - J_3 \right) \sum_\mu e^{i {\bf k}\cdot {\bf e}_\mu},\\
&& \gamma_{3,y}^{12} = \gamma_{3,y}^{13} = \gamma_{3,y}^{14} = \gamma_{3,y}^{2,3} = \gamma_{3,y}^{2,4} = \gamma_{3,y}^{3,4} = 0,
\end{eqnarray}
for fully characterizing $\mathcal{H}_{3,y}$. Similar, we have
\begin{eqnarray}
&&\alpha_{3,z}^{11} = \alpha_{3,z}^{22} = \alpha_{3,z}^{33} = \alpha_{3,z}^{44} = J_3 \sum_\mu \cos({\bf k}\cdot {\bf e}_1) - 2 K_3,\\
&& \alpha_{3,z}^{12} = \alpha_{3,z}^{13} = \alpha_{3,z}^{14} = \alpha_{3,z}^{23} = \alpha_{3,z}^{24}=\alpha_{3,z}^{34}=0,
\end{eqnarray}
and
\begin{eqnarray}
&& \gamma_{3,z}^{11} = \gamma_{3,z}^{22} =\gamma_{3,z}^{33} = \gamma_{3,z}^{44} = \left(K_3 - J_3\right) \sum_\mu e^{i {\bf k}\cdot {\bf e}_\mu},\\
&& \gamma_{3,z}^{12} = \gamma_{3,z}^{13} =\gamma_{3,z}^{14}=\gamma_{3,z}^{23} = \gamma_{3,z}^{24}= \gamma_{3,z}^{34} = 0
\end{eqnarray}
for fully characterizing $\mathcal{H}_{3,z}$. Combing the matrices obtains the Hermitian matrices used in the main text, $ \alpha_{\eta = y,z} = \sum_{n=1,2,3} \alpha_{n,\eta},$ and $\gamma_{\eta = y,z} = \sum_n \gamma_{n,\eta}.$

\clearpage
\section{DMRG: data in the real space for the $(\pi,0)$ AFQ phase and analyses for the finite-size scalings of the different phases}
\begin{figure}[h]
   \centering
   \includegraphics[width=3 in]{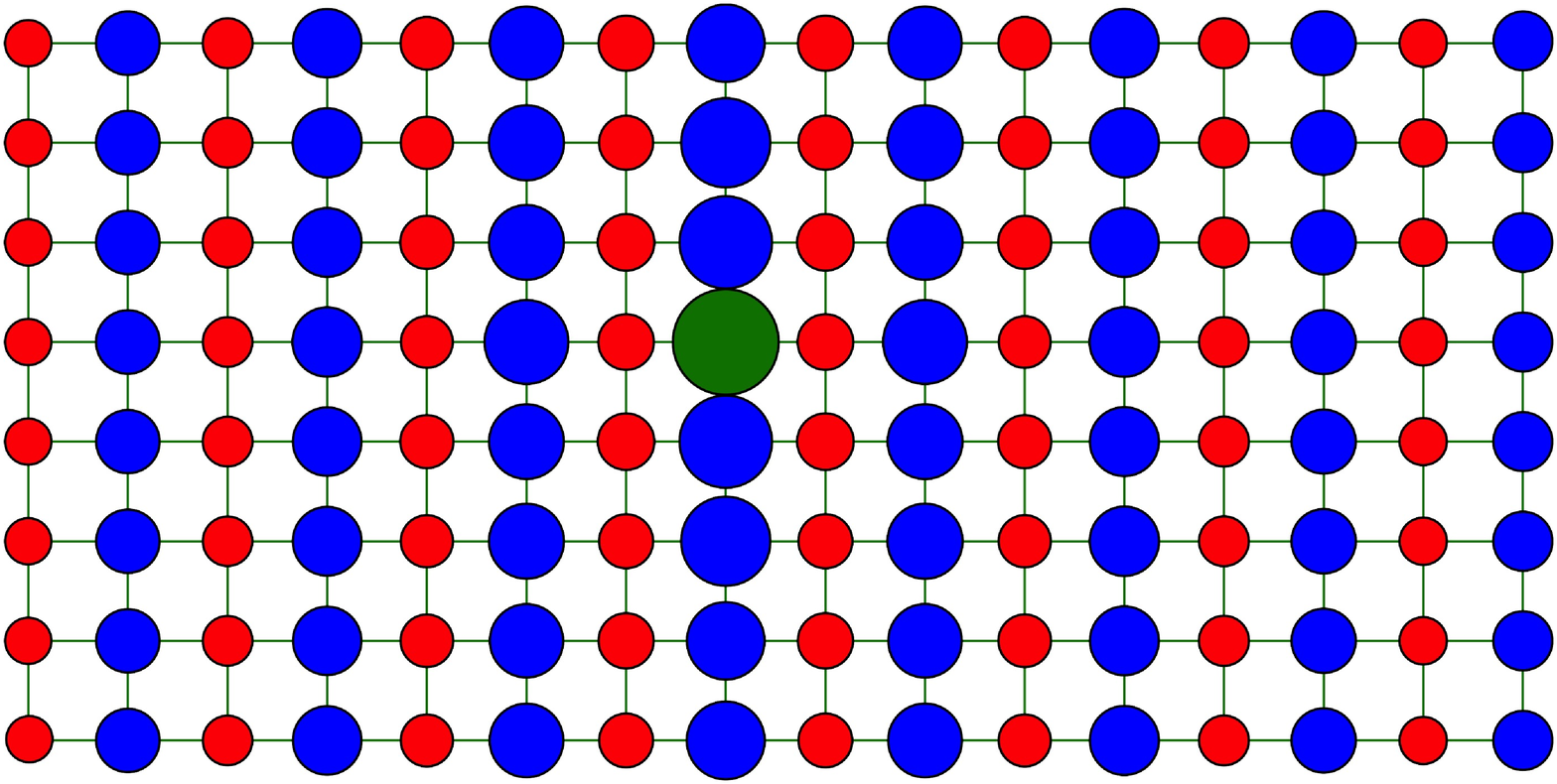}
   \caption{(Color online) The real-space quadrupolar correlation function on the $8\times16$ cylinder with $(K_2,K_3)=(1,-1)$. The green site is the reference site; the blue and red colors denote positive and negative correlations of the sites with the reference site, respectively. The area of circle is proportional to the magnitude of the quadrupolar correlation.}
   \label{Fig:afq}
\end{figure}

In this section we show the DMRG results of real-space quadrupolar correlation functions in the $(\pi,0)$ AFQ phase and the spin dipolar and quadrupolar structure factors in different phases at specified parameter points shown in the Fig. 4 in the main texts as a function of $1/L$ $(L_y = L)$. Fig.~\ref{Fig:afq} illustrates the quadrupolar correlation function in real space. We choose the reference site (green) in the middle of cylinder with $8\times16$, and compute the quadrupolar correlation. It clearly shows the staggered pattern along the $x$-direction, which indicates the $(\pi,0)$ AFQ order. 

\begin{figure}[h]
\centering
    \includegraphics[width= 3.5 in]{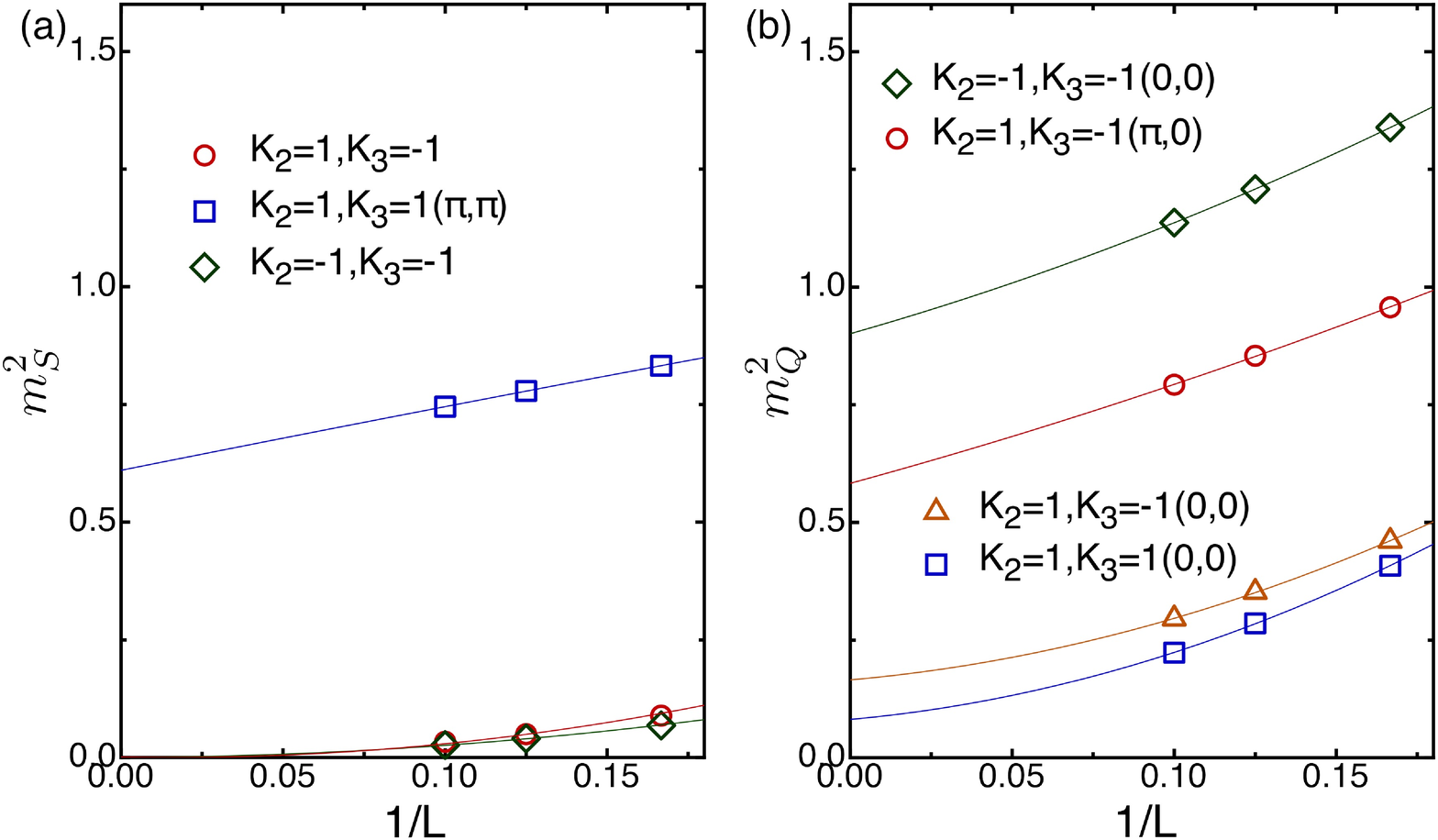}
    \caption{(Color online) The peak of the spin dipolar (a) and quadrupolar (b) structure factors, $m_s^2({\bf q})$ and $m_Q^2({\bf q})$, in different phases as a function of $1/L$ $(L_y = L)$. The lines are guides to the eye. The DMRG calculations have been done up to $L_y=10$ lattice $(J_1,J_2,J_3,K_1)=(1/4, 1/4, 0, -1)$ for N\'{e}el order with $(K_2,K_3) = (1,1)$ ), FQ with $(K_2,K_3) = (-1,-1$), and $(\pi,0)$ AFQ with $(K_2,K_3) = (1,-1)$.
    }
 \label{Fig:DMRG}
\end{figure}

Figs.~\ref{Fig:DMRG}(a)-(b) show $m_S^2({\bf q})$ and $m_Q^2({\bf q})$ at different wave vectors in different phases as a function of $1/L$. The blue open squares, green open diamonds, and red open circles (along with orange open triangles) represent, respectively, the data for $(K_2, K_3) = (1,1),~(-1,-1),$ and $(1,-1)$ with $(J_1,J_2,J_3,K_1)=(1/4, 1/4, 0, -1)$. At $(K_2, K_3) = (1,-1)$ the $m^2_{Q}(\pm{\pi},0)$ and $m^2_Q(0,0)$ peaks only slightly decrease from $L=6$ to $10$, while that of $m^2_{S}$ is further suppressed upon increasing $L$ to $10$, suggesting a vanishing $m_S^2$ upon further increasing $L$. The DMRG results provide evidence for the $(\pi,0)$ AFQ order. At $(K_2, K_3) = (1,1)$, $m_S^2(\pi,\pi)$ barely decreases upon increasing $L$ indicating this is the N\'{e}el AFM. At $(K_2, K_3) = (-1,-1)$, the strong $m_Q^2(0,0)$ peak along with the vanishing $m_s^2$ indicate that it is FQ.

\clearpage

\section{$(\pi,0)$ AFQ phase in the smaller biquadratic coupling regimes}
\begin{figure}[h]
\centering
    \includegraphics[width= 3.5 in]{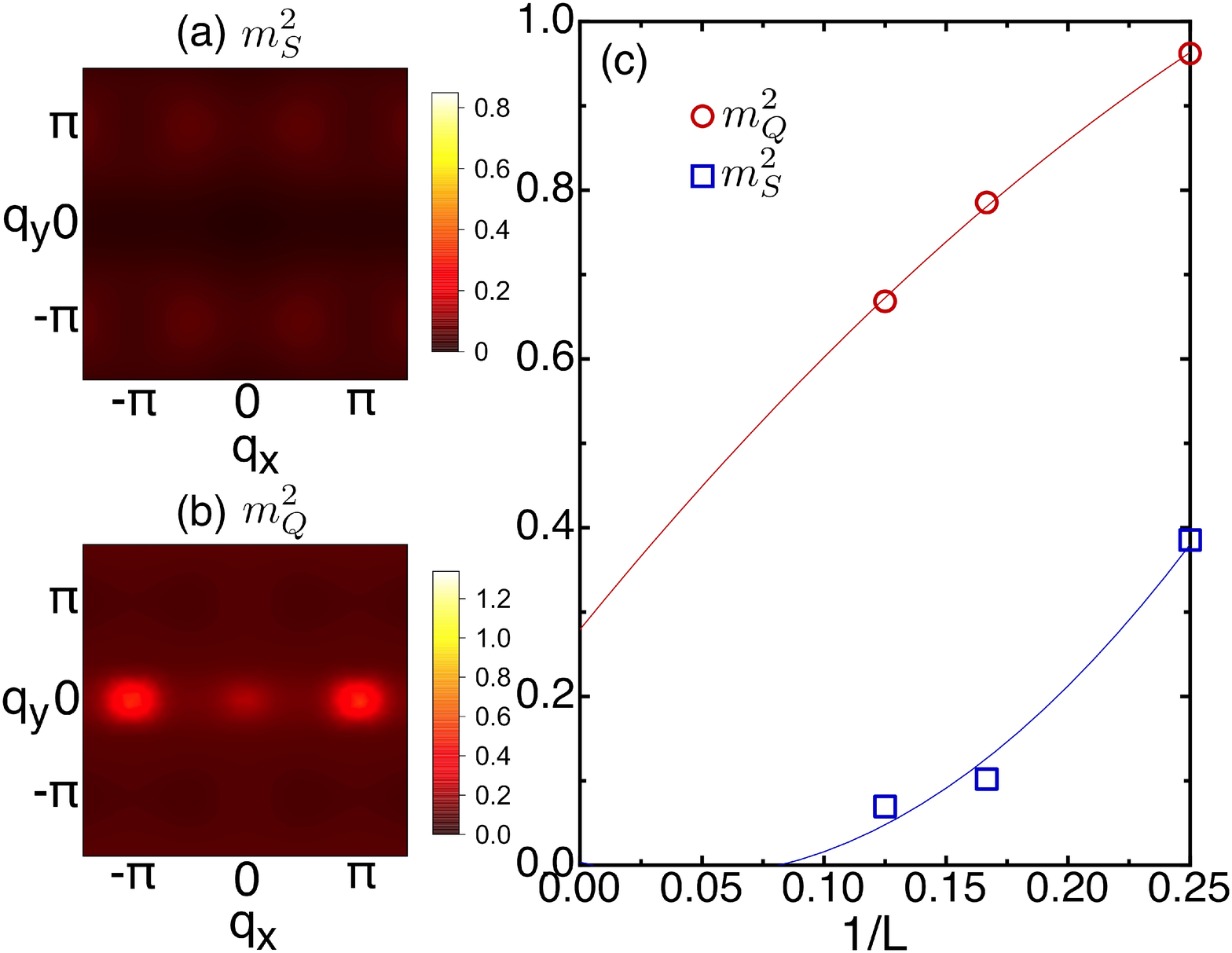}
    \caption{((Color online) The spin dipolar (a) and quadrupolar (b) structure factors, $m^2_S(q)$ and $m^2_Q(q)$, with
$(J_1,J_2,J_3,K_1,K_2,K_3)=(1/4,1/4,0,-1/5,1/5,-1/5)$ on $L_y=8$ lattice. (c) The finite-size scaling of the peaks of $m^2_S(q)$ and $m^2_Q(q)$ as a function of $1/L$ $(L_y=L)$. The lines are guides to the eye.
    }
 \label{Fig:DMRG_smallK}
\end{figure}

The $(\pi,0)$ AFQ phase does not require fine-tuning of parameters, and can be stable over a large parameter space. For an illustration, we fix $J_1,~J_2,~J_3$ to be the same as those considered in the main text and choose $-K_1 = K_2 = -K_3$ to be smaller than $J_2$. Specifically, we take $(J_1, J_2, J_3, K_1, K_2,K_3) = (1/4, 1/4, 0, -1/5, 1/5, -1/5)$. We show that the $(\pi,0)$ AFQ phase is still stable by observing the vanishing spin dipolar order and the nonzero quadrupolar order as shown in Fig.~\ref{Fig:DMRG_smallK}. Panels (a)-(b) show the spin dipolar structure factor $m_S^2(\bm{q})$ and quadrupolar structure factor $m_Q^2(\bm{q})$, respectively.  We can see vanishing peaks in Fig.~\ref{Fig:DMRG_smallK}(a) and, in Fig.~\ref{Fig:DMRG_smallK}(b), strong peaks at $(\pm \pi, 0)$ along with a weaker peak at $(0,0)$. Fig.~\ref{Fig:DMRG_smallK}(c) shows that the peak of the spin dipolar structure factor extrapolates to zero in the thermodynamic limit, while the peaks of the spin quadrupolar structure factor extrapolate to a nonzero value. These results are fully consistent with the $(\pi,0)$ AFQ ground state.

\clearpage

\section{Couplings between local moments and multi-orbital conduction electrons in the absence of spin-orbit couplings}

In this section we determine the lowest-order symmetry-allowed couplings between atomic quadrupole and conduction electron bilinear operators. In general, these are of the form

\noindent \begin{equation}
\label{Eq:Gnrl_frm}
H_{\text{int}}= I \sum_{\mu} \sum_{ijkl\dots} \sum_{\alpha \beta \gamma \delta \dots} Q_{\text{atom}}^{(\mu)} \left( c^{\dagger}_{\alpha, i} M_{\alpha \beta, ij}^{(\mu)} c_{\beta, j} \right) \left( c^{\dagger}_{\gamma, k} M_{\gamma \delta, kl}^{(\mu)} c_{\delta, l} \right) \dots ,
\end{equation}

\noindent where $\mu$ represents the five independent components of the atomic quadrupole, $\alpha, \beta, \gamma,\delta \dots$ are orbital indices and $i,j,k,l \dots$ are the spin-1/2 projection numbers for the conduction electrons. For simplicity, the analogous indices were suppressed for $Q_{\text{atom}}^{(\mu)}$. We consider only on-site couplings, although the generalization to non-local forms is straightforward. 

The $M$ matrices of Eq.~(\ref{Eq:Gnrl_frm}) are constrained by the rotation-symmetry group $G$ of the zero-coupling ($I=0$) Hamiltonian in Eq.~(1) of the main text. In the absence of spin-orbit coupling, the real-space rotations are independent of rotations in spin-space and the symmetry group $G$ is a product of the finite $D_{4h}$ point-group and the continuous SU(2) group for the two sectors respectively. In this case, $Q^{(\mu)}_{\text{atom}}$ acts only on the spin-1 atomic states and has no orbital dependence.  The $M$ matrices can be written as tensor products of matrices in orbital and spin-space:

\noindent \begin{align}
 M_{\alpha \beta, ij}=\tau_{\alpha \beta} \otimes \sigma_{ij}.
\end{align}

\noindent The orbital-matrix $\tau_{\alpha \beta}$ is in general a $5 \times 5$ matrix in the space of all $d_{xz}, d_{yz}, d_{xy}, d_{x^2-y^2}$, and $d_{z^2}$ orbitals. 

We show that the leading-order terms in $H_{\text{int}}$ which are invariant under $G$ involve at least two conduction-electron bilinear operators. This is done using two different methods. The first consists in explicitly constructing $H_{\text{int}}$ using irreducible tensors of the group $G$. The second relies on a generic coupling of SU(2) generators, which are also invariant under $D_{4h}$. The two approaches agree. Note that, in constructing the form of the quadrupole-bilinear couplings, we follow the common procedure and determine the coupling of the order-parameter field to itinerant electronic degrees-of-freedom in the disordered phase. When the order sets in, the order-parameter field acquires a static value and the form of the coupling will determine the effect of the order parameter on the itinerant electrons.
\subsection{I.~Irreducible tensors}

In this section, we construct SU(2)-invariant couplings between atomic spin-quadrupolar and fermion-bilinear operators using irreducible tensor operators of the rotation group $G$. Equivalently, we consider products of irreducible tensors which transform as the identity representation of $G$.

In general, for both finite groups and compact Lie groups the irreducible tensor operators $Q^{(q)}_{n}$ are defined as~\cite{Cornwell_1984_vol_1}

\noindent \begin{equation}
\label{Eq:Irrdcbl_tnsrs}
\Phi(T) Q^{(q)}_{n}=\sum^{d_{q}}_{m=1} \Gamma^{(q)}_{mn}(T)Q^{(q)}_{m},
\end{equation}

\noindent where the linear operator $\Phi(T)$ and the matrix $\Gamma(T)$ correspond to the element $T \in G$, while $d_{q}$ is the dimension of the representation $\Gamma^{(q)}$. For continuous Lie groups, an analogous definition in terms of Lie algebras can also be given in Ref.~\cite{Cornwell_1984_vol_2}. From the point of view of symmetry, the action of an irreducible tensor $Q^{(q)}_{p}$ on an arbitrary function is equivalent to multiplication by a basis function $\psi^{(q)}_{p}$ belonging to the $p$th row of the irreducible representation $\Gamma^{(q)}$~\cite{Cornwell_1984_vol_1}. The sought-after invariant can be determined by a Clebsch-Gordan series for the product of irreducible tensors of $G$.

In Table~\ref{Tbl:Mltpl_mmnts} we express the atomic spin-dipole and spin-quadrupole operators $Q^{(\mu)}$ in terms of irreducible tensors of SU(2)~\cite{Shiina_Shiba_Thalmeier:JPSJ_1997}. Note that the quadrupole operators decompose in terms of rank-2 irreducible tensors. As the latter transform as $J=2$ angular momentum states, we require analogous rank-2 irreducible operators in the conduction-electron sector in order to construct irreducible tensor states of rank 0, which are invariant under all SU(2) rotations. 

\begin{table}[t]
\caption{\label{Tbl:Mltpl_mmnts} Spin-multipole moments and their decomposition in terms of irreducible tensors $Q^{(q)}_{p}$ defined for SU(2).~\cite{Shiina_Shiba_Thalmeier:JPSJ_1997}}
\begin{tabularx}{1\textwidth}{*3{>{\centering\arraybackslash}X}}
\hline 
\hline 
Moment & Angular momentum representation & Decomposition into irreducible tensors  \\
\hline
Dipole & $J_{x}$ & $\frac{1}{\sqrt{2}} \left(-Q^{(1)}_{1} + Q^{(1)}_{-1} \right)$ \\
& $J_{y}$ & $\frac{1}{\sqrt{2}} \left(Q^{(1)}_{1} + Q^{(1)}_{-1} \right)$ \\
 & $J_{z}$ & $Q^{(1)}_{0}$ \\
\hline
Quadrupole & $Q_{3z^2-r^2}=\frac{1}{2}\left(2J^{2}_{z} - J^{2}_{x} - J^{2}_{y} \right)$ & $Q^{(2)}_{0}$ \\
& $Q_{x^2-y^2}=\frac{\sqrt{3}}{2} \left(J^{2}_{x} - J^{2}_{y} \right)$ & $\frac{1}{\sqrt{2}} \left(Q^{(2)}_{2} + Q^{(2)}_{-2} \right)$ \\
& $Q_{yz}=\frac{\sqrt{3}}{2} \left(J_{y}J_{z} + J_{z}J_{y}\right)$ & $\frac{i}{\sqrt{2}} \left(Q^{(2)}_{1} + Q^{(2)}_{-1} \right)$ \\
& $Q_{xz}=\frac{\sqrt{3}}{2} \left(J_{x}J_{z} + J_{z}J_{x}\right)$ & $\frac{i}{\sqrt{2}} \left(-Q^{(2)}_{1} + Q^{(2)}_{-1} \right)$ \\
& $Q_{xy}=\frac{\sqrt{3}}{2} \left(J_{x}J_{y} + J_{y}J_{x}\right)$ & $\frac{i}{\sqrt{2}} \left(-Q^{(2)}_{2} + Q^{(2)}_{-2} \right)$ \\
\hline
\hline
\end{tabularx}
\end{table}

By analogy with their atomic counterparts, we construct the local, orbital-diagonal, angular momentum density operators for the conduction electrons as

\noindent \begin{align}
& s_{x, \alpha} = \frac{1}{2} \sum_{ij} c^{\dagger}_{\alpha, i} (\sigma_{x})_{ij} c_{\alpha, j} = \frac{1}{\sqrt{2}} \left (-s^{(1)}_{1, \alpha} - s^{(1)}_{-1, \alpha} \right) \label{Eq:s_x_dfntn} \\
& s_{y, \alpha} = \frac{1}{2} \sum_{ij} c^{\dagger}_{\alpha, i} (\sigma_{y})_{ij} c_{\alpha, j} = \frac{1}{\sqrt{2}} \left (s^{(1)}_{1, \alpha} - s^{(1)}_{-1, \alpha} \right) \label{Eq:s_y_dfntn} \\
& s_{z, \alpha} = \frac{1}{2} \sum_{ij} c^{\dagger}_{\alpha, i} (\sigma_{z})_{ij} c_{\alpha, j} = s^{(1)}_{0}, \label{Eq:s_z_dfntn} 
\end{align}

\noindent where $\alpha$ is an orbital index $\alpha \in \{d_{xz}, d_{yz}, d_{xy}, d_{x^2-y^2}, d_{3z^2-r^2}\}$. For simplicity, we restrict ourselves to the two-orbital ${d_{xz}, d_{yz}}$ sector as the extension to larger orbital spaces is straightforward. In order to get invariance under $D_{4h}$ we define orbital-independent fermion-bilinear operators  as

\noindent \begin{equation}
\label{Eq:Ttl_spn_oprtrs}
s_{k}= \sum_{\alpha \in \{d_{xz}, d_{yz}\}} s^{(1)}_{k, \alpha},~~~ k \in \{x, y, z\}. 
\end{equation}

\noindent For rotations of the point-group $D_{4h}$ these operators transform like the identity representation, or equivalently as $d_{xz}d_{xz}+ d_{yz}d_{yz}$. Note that the operator $\pmb{s}=\sum_{k \in \{x, y, z\}} s_{k, d_{xz}} + s_{k, d_{yz}}$ can also be interpreted as a total spin operator on the product space of single-particle, local,  $d_{xz}$ and $d_{yz}$ conduction-electron states. These product states decompose into a singlet space of $s_{tot}=0$ and a triplet of $s_{tot}=1$. Therefore, we can construct quadrupole operators, or equivalently, rank-2 irreducible tensors in the local conduction-electron product states. In the absence of two different orbitals, or equivalently, in a spin-1/2 representation, all these quadrupole operators would vanish. Consequently, we can define rank-1 and rank-2 irreducible operators for the conduction-electron states in direct analogy with the atomic states. These will be denoted by lowercase $q^{(p)}_{q}$. 

The last step consists in constructing products of irreducible tensors which transform as the identity representation of SU(2). This is equivalent to constructing a $J=0$ angular momentum state out of $J_{1}=2$ and $J_{2}=2$ states. The Clebsch-Gordan coefficients are readily available and the resulting SU(2)-invariant coupling can be written as

\noindent \begin{equation}
\label{Eq:Invrnt_frm}
Q^{(0)}_{0}=\frac{1}{5} \left[ Q^{(2)}_{2} q^{(2)}_{-2} - Q^{(1)}_{1}q^{(2)}_{-1} + Q^{(2)}_{0}q^{(2)}_{0} -  Q^{(1)}_{-1}q^{(2)}_{1} + Q^{(2)}_{-2} q^{(2)}_{2} \right].
\end{equation}
 
\noindent Expressing the irreducible tensors in terms of the atomic multi-pole operators $\pmb{J}$, $\pmb{Q}$, and their conduction-electron counterparts $\pmb{s}$, $\pmb{q}$, Eq.~(\ref{Eq:Invrnt_frm}) becomes  

\noindent \begin{equation}
\label{Eq:Qdrpl-qdrpl}
H_{\text{int}}=Q_{z^2} \cdot q_{z^2} + Q_{xz} \cdot q_{xz} + Q_{yz} \cdot q_{yz} + Q_{xy} \cdot q_{xy} + Q_{x^2-y^2} \cdot q_{x^2-y^2}=\pmb{Q} \cdot \pmb{q}.
\end{equation}

\noindent The explicit construction of $H_{\text{int}}$ shows that conduction electrons can only couple at the biquadratic or higher orders for $G=D_{4h} \times $ SU(2). 

\subsection{II.~Lie-algebra approach}

In this subsection, we reproduce the results and conclusion of the previous subsection using a more common approach based on Lie algebras. Indeed, the conduction-electron bilinear operators $\pmb{s}$ defined in Eqs.~(\ref{Eq:s_x_dfntn})-(\ref{Eq:Ttl_spn_oprtrs}), form an irreducible representation of the SU(2) algebra in the total spin conduction electron space. As explained by previous arguments, they are also invariant with respect to $D_{4h}$. Consider a coupling of the form

\noindent \begin{align}
H_{\text{dipole}}= & L \pmb{J} \cdot \pmb{s} \notag \\
= & L \left( C_{2}(\pmb{J} + \pmb{s}) - C_{2}(\pmb{J}) - C_{2}(\pmb{s}) \right).
\end{align}

\noindent In the above expression, $\pmb{J}$ stands for atomic dipoles, and $C_{2}(X)$ stand for second-order Casimir operators~\cite{Cornwell_1984_vol_2} which commute with all the elements $X$ of an SU(2) algebra. In other words, $H_{\text{dipole}}$ is invariant under simultaneous rotations generated by the total angular momentum operator $\pmb{J} + \pmb{s}$, and is thus SU(2)-invariant. It is also invariant under $D_{4h}$, and thus under $G$. 

While $H_{\text{dipole}}$ is $G$-invariant, it does not contain atomic quadrupole degrees-of-freedom. In order to generate these, we introduce higher-order couplings given by

\begin{align}
H_{\text{quadrupole}}=& K (\pmb{S} \cdot \pmb{s})^2 \\
= &  K \left[\frac{1}{2} \pmb{Q} \cdot \pmb{q} -\frac{1}{2} \pmb{S} \cdot \pmb{s} + \frac{1}{3}\pmb{S}^{2} \pmb{s}^{2} \right].
\end{align}

\noindent In the second line we used the expressions for the atomic and conduction electron dipoles and quadrupoles in terms of the angular momentum operators as given by Table~\ref{Tbl:Mltpl_mmnts}. As $H_{\text{quadrupole}}$ together with the second and third terms on the second line are all SU(2) and $G$-invariant, we conclude that the $\pmb{Q} \cdot \pmb{q}$ terms are also SU(2) and $G$ invariant. These are the same terms in $H_{\text{int}}$ obtained in the previous subsection. Based on the very general procedure used to obtain $H_{\text{quadrupole}}$, we conclude that the lowest-order couplings between the atomic quadrupoles and the conduction electron bilinears involve at least two of the former. 

\section{Further extension of the model}
In this work, we have focused on a model with only nearest- and next-nearest-neighbor bilinear couplings. Adding the third-neighbor bilinear coupling will further extend the phase diagram shown in Fig.~1 in the main texts. The phase diagram, and its extension to nonzero temperatures, provides the basis to understand the richness in the magnetic and related spin orders and their phase transitions in FeSe, both at ambient condition and under pressure~\cite{Bendele10,Bendele12,Imai09}, in the  FeSe$_{1-x}$Te$_x$ series~\cite{Xu2015} and, more generally, in the context of comparing FeSe with other FeSCs.